\documentclass[fleqn,usenatbib]{mnras}

\usepackage{newtxtext,newtxmath}

\usepackage[T1]{fontenc}

\DeclareRobustCommand{\VAN}[3]{#2}
\let\VANthebibliography\thebibliography
\def\thebibliography{\DeclareRobustCommand{\VAN}[3]{##3}\VANthebibliography}

\usepackage{graphicx}	
\usepackage{amsmath}	
\usepackage{subcaption}
\usepackage{dsfont}
\usepackage{booktabs}
\usepackage{orcidlink}
\usepackage{widetext}
\usepackage[dvipsnames]{xcolor}
\usepackage[colorinlistoftodos, color=lime]{todonotes}

\DeclareMathAlphabet{\mathcal}{OMS}{cmsy}{m}{n} 

\newcommand{\bold}[1]{\boldsymbol{\mathrm{#1}}}

\title[Time Series Calibration]{Capturing System Drift with Time Series Calibration for Global 21-cm Cosmology Experiments}

\author[C. J. Kirkham et al.]{
Christian J. Kirkham \textsuperscript{\orcidlink{0000-0001-5385-6329}},$^{1,2}$\thanks{E-mail: \href{mailto:cjk55@cam.ac.uk}{cjk55@cam.ac.uk}} 
Dominic J. Anstey \textsuperscript{\orcidlink{0000-0003-1742-7417}},$^{1,2}$\thanks{E-mail: \href{mailto:da401@cam.ac.uk}{da401@cam.ac.uk}} and Eloy de Lera Acedo \textsuperscript{\orcidlink{0000-0001-8530-6989}}$^{1,2}$\thanks{E-mail: \href{mailto:ed330cam.ac.uk}{ed330@cam.ac.uk}}
\\
$^{1}$Astrophysics Group, Cavendish Laboratory, J. J. Thomson Avenue, Cambridge, CB3 0HE, UK\\
$^{2}$Kavli Institute for Cosmology, Madingley Road, Cambridge, CB3 0HA, UK\\
}

\date{Accepted XXX. Received YYY; in original form ZZZ}

\pubyear{2025}

\begin{document}
\label{firstpage}
\pagerange{\pageref{firstpage}--\pageref{lastpage}}
\maketitle

\begin{abstract}
To achieve the sensitivity required to detect signals from neutral hydrogen from the Cosmic Dawn and Epoch of Reionisation it is critical to have a well-calibrated instrument which has a stable calibration over the course of the observation. Previous Bayesian calibration methods do not explicitly use the time information available and make assumptions on the impedance matching of the reference sources. Here we present a new calibration method based on noise wave parameters which fits a calibration solution over time and frequency to the data, interpolating the solutions to the times at which the antenna is being measured. To test this method we simulate a dataset using measurements of the REACH receiver, modelling a low noise amplifier which is drifting over time. Fitting a polynomial surface in frequency and time to the simulated data demonstrates that we can remove the drift in the calibrated solution over time but leaves a chromatic residual. We further show that we can remove assumptions on the reflection coefficients of the reference noise source and the cold load, reducing degeneracies in the parameter fits. Applying this new calibration equation and surface fitting method to the simulated data removes the chromatic residual in the calibrated spectrum and recovers the parameters to within 0.06\% of the truth and a 97\% reduction in the RMSE of the spectrum of the validation source compared with previous calibration methods. For two parameters we report up to six times smaller fit error after the degeneracies are removed from the time-based calibration.
\end{abstract}




\section{Introduction}

21-cm radio cosmology experiments aim to infer the properties of the first stars and galaxies from radio emissions from neutral hydrogen (HI) in the Cosmic
Dark Ages, Cosmic Dawn, and the Epoch of Reionization \citep{furlanettoCosmologyLowFrequencies2006, monsalveResultsEDGESHighband2018, monsalveResultsEDGESHighBand2019, bevinsAstrophysicalConstraintsSARAS2022}. They do so by measuring the redshifted hyperfine transition line which is at a rest wavelength of $\lambda \approx 21\,$cm or a frequency of $\nu \approx 1420\,$MHz as a statistical `spin temperature', measured relative to the temperature of cosmic microwave background.

Global 21-cm radio experiments use low-frequency radio antennae to detect this radio emission from the neutral hydrogen. Examples of such experiments are the Experiment to Detect the Global EoR Signature (EDGES) \citep{bowmanEmpiricalConstraintsGlobal2008}, Shaped Antenna measurement of the background Radio Spectrum (SARAS) \citep{singhSARASSpectralRadiometer2018, singhDetectionCosmicDawn2022}, Large Aperture Experiment to Detect the Dark Ages (LEDA) \citep{priceDesignCharacterizationLargeaperture2018}, Probing Radio Intensity at high-Z from Marion (PRI$^\mathrm{Z}$M) \citep{philipProbingRadioIntensity2019}, Mapper of the IGM Spin Temperature (MIST) \citep{monsalveMapperIGMSpin2023} and Radio Experiment for the Analysis of Cosmic Hydrogen (REACH) \citep{deleraacedoREACHRadiometerDetecting2022a}.

The first experiment to make a claimed detection of the 21-cm signal was the EDGES experiment \citep{bowmanAbsorptionProfileCentred2018} who found a flattened Gaussian absorption profile of depth $500^{+500}_{-200}\,$mK centred at $78\pm1\,$MHz. The unusual depth and shape of this signal however requires exotic physics to explain it \citep{barkanaPossibleInteractionBaryons2018, fengEnhancedGlobalSignal2018}. Recent observations by the SARAS experiment have placed constraints on this signal, ruling out the EDGES detection with 95.3\% confidence \citep{singhDetectionCosmicDawn2022}.

Concerns have been raised that there is an unmodelled systematic in the EDGES data \citep{hillsConcernsModellingEDGES2018, simsTestingCalibrationSystematics2020} which is causing a biased fit. \cite{hillsConcernsModellingEDGES2018} found that comparing the EDGES foreground fit to a physical model resulted in unphysical parameters, noting that the fit improves with the inclusion of a $12.5\,$MHz sinusoid. The presence of this sinusoid is supported by the work of \cite{simsTestingCalibrationSystematics2020}, who note that this could be the result of a calibration error.


As the global 21-cm signal absorption depth is predicted to be small -- with an amplitude $\lesssim200\,$mK \citep{dhandhaNarrowingDiscoverySpace2025} -- while the free-free and synchrotron emission from the galactic foregrounds will be of order $10^3-10^4\,$K \citep{shaverCanReionizationEpoch1999}, we require highly sensitive experiments to separate the signal from the foregrounds. To do so requires long integration times, with a REACH-like experiment requiring approximately 6.5 hours of integration at minimum to reduce the radiometer noise down to the level at which it can detect the EDGES signal \citep{deleraacedoREACHRadiometerDetecting2022a}. As a result of this long integration time, we need to maintain a stable calibration over the observation to achieve the sensitivity necessary to detect the 21-cm signal. Previous calibration methods \citep{rogersAbsoluteCalibrationWideband2012, roqueBayesianNoiseWave2021} have not considered calibration over time and so require the instrument to be kept stable over the course of the observation.

This work will present a new method of calibration for global 21-cm cosmology experiments which fits a frequency-time surface to the calibration parameters, interpolating to the antenna observations between calibrator observations. This work will focus on calibrating the REACH experiment \citep{deleraacedoREACHRadiometerDetecting2022a}, introducing the calibration of the receiver system in section \ref{s:methods-receiver}. In section \ref{s:methods-assumptions} we will introduce a new way of calibrating a receiver to remove parameter degeneracies and in section \ref{s:methods-surface} we present the surface fitting method. Section \ref{s:methods-simulation} outlines the generation of simulated data to test the method and in section \ref{s:results} we present the results. Finally in section \ref{s:conclusions} we present the conclusions.

\section{Methods}

\subsection{REACH Receiver Calibration} \label{s:methods-receiver}

Following \cite{rogersAbsoluteCalibrationWideband2012} and \cite{monsalveCALIBRATIONEDGESHIGHBAND2017}, calibration of the REACH receiver is done using a combination of Dicke switching \citep{dickeMeasurementThermalRadiation1946} and noise wave parameters (NWP) introduced by \cite{meysWaveApproachNoise1978}. In REACH the Dicke switching is done by switching between a reference source (or the antenna), a noise source with noise temperature $T_\mathrm{NS} \approx 1400\,$K, and a cold load at ambient temperature $T_\mathrm{L}$ which are theoretically matched to the low noise amplifier (LNA). Using this noise wave parametrisation we can write the calibrated temperature in terms of the measured power spectral densities (PSD), $P$, and the reflection coefficients, $\Gamma$, \citep{roqueBayesianNoiseWave2021}
\begin{align} \label{e:full_calibration_eqn}
\begin{split}
    T'_\mathrm{NS} \left(\frac{P_\mathrm{s} - P_\mathrm{L}}{P_\mathrm{NS} - P_\mathrm{L}}\right) + T'_\mathrm{L} = &T_\mathrm{s} \left[ \frac{1 - |\Gamma_\mathrm{s}|^2}{|1 - \Gamma_\mathrm{s} \Gamma_\mathrm{r}|^2} \right]\\
    &+ T_\mathrm{unc} \left[ \frac{|\Gamma_\mathrm{s}|^2}{|1 - \Gamma_\mathrm{s} \Gamma_\mathrm{r}|^2} \right]\\
    &+ T_\mathrm{cos} \left[ \frac{\mathrm{Re}\left( \frac{\Gamma_\mathrm{s}}{1 - \Gamma_\mathrm{s}\Gamma_\mathrm{r}} \right)}{\sqrt{1 - |\Gamma_\mathrm{r}|^2}} \right]\\
    &+ T_\mathrm{sin} \left[ \frac{\mathrm{Im}\left( \frac{\Gamma_\mathrm{s}}{1 - \Gamma_\mathrm{s}\Gamma_\mathrm{r}} \right)}{\sqrt{1 - |\Gamma_\mathrm{r}|^2}} \right],
\end{split}
\end{align}
where $T_\mathrm{unc}$, $T_\mathrm{cos}$, $T_\mathrm{sin}$ are the frequency-dependent `noise wave parameters' which are unknowns and are fitted for. Here $P_\mathrm{s}$ and $\Gamma_\mathrm{s}$ are the source received power and reflection coefficient respectively, $\Gamma_\mathrm{r}$ is the LNA reflection coefficient, $P_\mathrm{NS}$ is the power received from the noise source and $P_\mathrm{L}$ is the cold load power. We measure the power with a spectrometer and the reflection coefficients with a vector network analyser (VNA), both of which are in the REACH receiver allowing full in-situ measurements of the system. Here we also fit for an effective noise source temperature, $T'_\mathrm{NS}$, and an effective cold load temperature, $T'_\mathrm{L}$, as frequency dependent parameters to account for any impedance mismatches between the LNA, the noise source and the cold load. Note that in the ideal case where there are no impedance mismatches, the effective noise source temperature has been defined such that $T'_\mathrm{NS} = T_\mathrm{NS} - T_\mathrm{L}$.

To constrain the five noise wave parameters the REACH receiver uses 12 reference calibration sources with a variety of temperatures and reflection coefficients to sample the parameter space. These calibration sources -- which have known temperatures measured in-situ -- are as follows:
\begin{itemize}
    \item An ambient 50$\,\Omega$ `cold' load
    \item Ambient 25$\,\Omega$ and 100$\,\Omega$ loads
    \item A 50$\,\Omega$ heated `hot' load at $370\,$K which is connected to a $4\,$in cable
    \item 27$\,\Omega$, 36$\,\Omega$, 69$\,\Omega$ and 91$\,\Omega$ ambient loads which are connected to a `short' $2\,$m cable (SC)
    \item 10$\,\Omega$, 250$\,\Omega$, open and short loads which are connected to a `long' $10\,$m cable (LC)
\end{itemize}
The antenna is a hexagonal dipole \citep{cumnerRadioAntennaDesign2022} which is connected to a cable approximately one metre in length.

We can rewrite equation \ref{e:full_calibration_eqn} by first defining the following $X$ coefficients,
\begin{equation} \label{e:x_matrices}
X_\mathrm{unc} = - \frac{|\Gamma_\mathrm{s}|^2}{1 - |\Gamma_\mathrm{s}|^2},
\end{equation}
\begin{equation} \label{e:X_load}
X_\mathrm{L} = \frac{|1 - \Gamma_\mathrm{s} \Gamma_\mathrm{r}|^2}{1 - |\Gamma_\mathrm{s}|^2},
\end{equation}
\begin{equation}
X_\mathrm{cos} = -\mathrm{Re}\left(\frac{\Gamma_\mathrm{s} }{1 - \Gamma_\mathrm{s}\Gamma_\mathrm{r}} \times \frac{X_\mathrm{L}}{\sqrt{1 - |\Gamma_\mathrm{r}|^2}}\right),
\end{equation}
\begin{equation}
X_\mathrm{sin} = -\mathrm{Im}\left(\frac{\Gamma_\mathrm{s} }{1 - \Gamma_\mathrm{s}\Gamma_\mathrm{r}} \times \frac{X_\mathrm{L}}{\sqrt{1 - |\Gamma_\mathrm{r}|^2}}\right),
\end{equation}
\begin{equation} \label{e:X_noise_source}
    X_\mathrm{NS} = \left(\frac{P_\mathrm{s} - P_\mathrm{L}}{P_\mathrm{NS} - P_\mathrm{L}}\right) X_\mathrm{L}.
\end{equation}
 These $X$ coefficients contain all of the quantities which are measured in-situ and allow us to rearrange equation \ref{e:full_calibration_eqn} into the linear equation
\begin{equation} \label{e:simplified_calibration_eqn}
    T_s(\nu) = X_\mathrm{unc} T_\mathrm{unc} + X_\mathrm{cos} T_\mathrm{cos} + X_\mathrm{sin} T_\mathrm{sin} + X_\mathrm{NS} T_\mathrm{NS}' + X_\mathrm{L} T_\mathrm{L}'.
\end{equation}

Finally we can define
\begin{equation}
    \bold{X} = \begin{pmatrix}X_\mathrm{unc} & X_\mathrm{cos} & X_\mathrm{sin} & X_\mathrm{NS} & X_\mathrm{L}\end{pmatrix},
\end{equation}
\begin{equation}
    \bold{\Theta} = \begin{pmatrix}T_\mathrm{unc} & T_\mathrm{cos} & T_\mathrm{sin} & T_\mathrm{NS}' & T_\mathrm{L}'\end{pmatrix}^T,
\end{equation}
which reduces the calibration equation down to the linear equation,
\begin{equation} \label{e:calibration_equation}
    \bold{T}_\mathrm{s} = \bold{X}\bold{\Theta},
\end{equation}
where $\bold{T}_\mathrm{s}$ is a vector of calibrated source temperatures over frequency.

To mitigate issues with varying calibrator noise and singularities we will fit the data using the $\Gamma$-weighted method of \cite{kirkhamAccountingNoiseSingularities2025}, where
\begin{equation}
    \tilde{\bold{T}} = (1 - |\Gamma_s|^2)\bold{T} = (1 - |\Gamma_s|^2)\bold{X}\bold{\Theta} = \tilde{\bold{X}}\bold{\Theta}.
\end{equation}

In REACH these parameters are fitted for using a variety methods such as the conjugate priors polynomial method \citep{roqueBayesianNoiseWave2021}, the least squares frequency-by-frequency method \citep{roqueReceiverDesignREACH2025} or the marginalised polynomial method \citep{kirkhamAccountingNoiseSingularities2025}. Alternative parametrisations used by REACH include the noise parameter formalisation of \cite{priceMeasuringNoiseParameters2023} which can be fitted using machine learning methods \citep{leeneyRadiometerCalibrationUsing2025a}. In this work we will introduce a new conjugate priors surface fitting method using the noise wave parametrisation described in this section.

\subsection{Removing Impedance Matching Assumptions and Degeneracies} \label{s:methods-assumptions}

Equation \ref{e:full_calibration_eqn} makes the assumption that the noise source and cold load are impedance matched to 50$\,\Omega$, i.e. $\Gamma_\mathrm{NS} = \Gamma_\mathrm{L} = 0$. In reality, where the impedance matching is imperfect, \cite{roqueBayesianNoiseWave2021} accounts for the mismatch by fitting frequency-dependent functions for $T_\mathrm{NS}'$ and $T_\mathrm{L}'$ which capture the realistic system\footnote{Similarly, in the EDGES calibration this is accounted for by introducing two parameters $C_1$ and $C_2$ \citep{monsalveCALIBRATIONEDGESHIGHBAND2017}. These two parameters also fit for the path difference between their noise source and the antenna.}. However, this introduces degeneracies into your noise wave parameter fit as it can be shown that

\begin{align}
\begin{split}
    T_\mathrm{NS}' = &\frac{1}{1-|\Gamma_\mathrm{rec}|^2}\Big[T_{\mathrm{NS}}
    \bigl(1 - \bigl|\Gamma_{\mathrm{NS}}\bigr|^{2}\bigr) |F_\mathrm{NS}|^2 - T_{\mathrm{L}}
    \bigl(1 - \bigl|\Gamma_{\mathrm{L}}\bigr|^{2}\bigr) |F_\mathrm{L}|^2\\
    &+
    T_{\mathrm{unc}} \left(
    \bigl|\Gamma_{\mathrm{NS}}\bigr|^{2} |F_\mathrm{NS}|^2 - \bigl|\Gamma_{\mathrm{L}}\bigr|^{2} |F_\mathrm{L}|^2\right) \\
    &+
    T_{\mathrm{cos}}
    \left(
    \bigl|\Gamma_{\mathrm{NS}}\bigr| |F_\mathrm{NS}| \cos\alpha_\mathrm{NS} - \bigl|\Gamma_{\mathrm{L}}\bigr| |F_\mathrm{L}|\cos\alpha_\mathrm{L}\right)\\
    &+
    T_{\mathrm{sin}}
    \left(
    \bigl|\Gamma_{\mathrm{NS}}\bigr| |F_\mathrm{NS}| \sin\alpha_\mathrm{NS} - \bigl|\Gamma_{\mathrm{L}}\bigr| |F_\mathrm{L}|\sin\alpha_\mathrm{L}\right)\Big],
\end{split}
\end{align}
and
\begin{align}
\begin{split}
    T_\mathrm{L}' = &\frac{1}{1-|\Gamma_\mathrm{rec}|^2}\Big[T_{\mathrm{L}}
    \bigl(1 - \bigl|\Gamma_{\mathrm{L}}\bigr|^{2}\bigr) |F_\mathrm{L}|^2 \\
    &+
    T_{\mathrm{unc}} \bigl|\Gamma_{\mathrm{L}}\bigr|^{2} |F_\mathrm{L}|^2 \\
    &+
    T_{\mathrm{cos}}
    \bigl|\Gamma_{\mathrm{L}}\bigr| |F_\mathrm{L}| \cos\alpha_\mathrm{L}\\
    &+
    T_{\mathrm{sin}}
    \bigl|\Gamma_{\mathrm{L}}\bigr| |F_\mathrm{L}| \sin\alpha_\mathrm{L}\Big],
\end{split}
\end{align}
which have dependencies on the other fitted noise wave parameters, $T_\mathrm{unc}$, $T_\mathrm{cos}$ and $T_\mathrm{sin}$. Here,
\begin{equation}
    F_i = \frac{\sqrt{1 - |\Gamma_\mathrm{rec}|^2}}{1- \Gamma_\mathrm{i}\Gamma_\mathrm{rec}},
\end{equation}
and
\begin{equation}
    \alpha_i = \mathrm{arg}(\Gamma_iF_i).
\end{equation}

We can remove these degeneracies and assumptions by rewriting the calibration equation to include the reflection coefficients of the noise source and cold load and rearranging into the linear form
\begin{equation}\label{e:calibration_eqn_TNSTL}
    T_s(\nu) = \bar{X}_\mathrm{unc}T_\mathrm{unc} + \bar{X}_\mathrm{cos}T_\mathrm{cos} + \bar{X}_\mathrm{sin}T_\mathrm{sin} + \bar{X}_\mathrm{NS}T_\mathrm{NS} + \bar{X}_\mathrm{L}T_\mathrm{L},
\end{equation}
where
\begin{equation}
    \bar{X}_\mathrm{NS} = \frac{P_\mathrm{s} - P_\mathrm{L}}{P_\mathrm{NS} - P_\mathrm{L}} \frac{(1 - |\Gamma_\mathrm{NS}|^2)|1 - \Gamma_\mathrm{s}\Gamma_\mathrm{rec}|^2}{(1 - |\Gamma_\mathrm{s}|^2)|1 - \Gamma_\mathrm{NS}\Gamma_\mathrm{rec}|^2},
\end{equation}
\begin{equation}
    \bar{X}_\mathrm{L} = \left[1 - \frac{P_\mathrm{s} - P_\mathrm{L}}{P_\mathrm{NS} - P_\mathrm{L}} \right] \frac{(1 - |\Gamma_\mathrm{L}|^2)|1 - \Gamma_\mathrm{s}\Gamma_\mathrm{rec}|^2}{(1 - |\Gamma_\mathrm{s}|^2)|1 - \Gamma_\mathrm{L}\Gamma_\mathrm{rec}|^2},
\end{equation}
\begin{equation}
    \bar{X}_\mathrm{unc} = \bar{X}_\mathrm{NS} \frac{|\Gamma_\mathrm{NS}|^2}{1 - |\Gamma_\mathrm{NS}|^2} + \bar{X}_\mathrm{L} \frac{|\Gamma_\mathrm{L}|^2}{1 - |\Gamma_\mathrm{L}|^2} - \frac{|\Gamma_\mathrm{s}|^2}{1 - |\Gamma_\mathrm{s}|^2},
\end{equation}
\begin{align}
\begin{split}
    \bar{X}_\mathrm{cos} &= \bar{X}_\mathrm{NS} \cdot \mathrm{Re} \left( \frac{\Gamma_\mathrm{NS}}{1 - \Gamma_\mathrm{NS}\Gamma_\mathrm{rec}}\cdot\frac{|1-\Gamma_\mathrm{NS}\Gamma_\mathrm{rec}|^2}{\sqrt{1 - |\Gamma_\mathrm{rec}|^2}(1 - |\Gamma_\mathrm{NS}|^2)} \right)\\
    &+\bar{X}_\mathrm{L} \cdot \mathrm{Re} \left( \frac{\Gamma_\mathrm{L}}{1 - \Gamma_\mathrm{L}\Gamma_\mathrm{rec}}\cdot\frac{|1-\Gamma_\mathrm{L}\Gamma_\mathrm{rec}|^2}{\sqrt{1 - |\Gamma_\mathrm{rec}|^2}(1 - |\Gamma_\mathrm{L}|^2)} \right)\\
    &-\mathrm{Re} \left( \frac{\Gamma_\mathrm{s}}{1 - \Gamma_\mathrm{s}\Gamma_\mathrm{rec}}\cdot\frac{|1-\Gamma_\mathrm{s}\Gamma_\mathrm{rec}|^2}{\sqrt{1 - |\Gamma_\mathrm{rec}|^2}(1 - |\Gamma_\mathrm{s}|^2)} \right),
\end{split}
\end{align}
\begin{align}
\begin{split}
    \bar{X}_\mathrm{sin} &= \bar{X}_\mathrm{NS} \cdot \mathrm{Im} \left( \frac{\Gamma_\mathrm{NS}}{1 - \Gamma_\mathrm{NS}\Gamma_\mathrm{rec}}\cdot\frac{|1-\Gamma_\mathrm{NS}\Gamma_\mathrm{rec}|^2}{\sqrt{1 - |\Gamma_\mathrm{rec}|^2}(1 - |\Gamma_\mathrm{NS}|^2)} \right)\\
    &+\bar{X}_\mathrm{L} \cdot \mathrm{Im} \left( \frac{\Gamma_\mathrm{L}}{1 - \Gamma_\mathrm{L}\Gamma_\mathrm{rec}}\cdot\frac{|1-\Gamma_\mathrm{L}\Gamma_\mathrm{rec}|^2}{\sqrt{1 - |\Gamma_\mathrm{rec}|^2}(1 - |\Gamma_\mathrm{L}|^2)} \right)\\
    &-\mathrm{Im} \left( \frac{\Gamma_\mathrm{s}}{1 - \Gamma_\mathrm{s}\Gamma_\mathrm{rec}}\cdot\frac{|1-\Gamma_\mathrm{s}\Gamma_\mathrm{rec}|^2}{\sqrt{1 - |\Gamma_\mathrm{rec}|^2}(1 - |\Gamma_\mathrm{s}|^2)} \right).
\end{split}
\end{align}
Note that with this formulation of the calibration equation $T_\mathrm{NS}$ and $T_\mathrm{L}$ are the noise temperature of the noise source and the physical temperature of the cold load respectively. Since we measure the ambient temperature throughout an observation it is no longer necessary to fit for $T_\mathrm{L}$. However, the datasheet value of $T_\mathrm{NS}$ is not accurate enough for receiver calibration so we will continue to fit for this as a parameter.

\subsection{Bayesian Inference}

To determine the noise wave parameters and the calibrated temperature noise, we employ Bayesian inference—a statistical framework that updates probabilities based on observed data using Bayes' theorem:
\begin{equation}
    P(\bold \theta|\bold D, \mathcal M) = \frac{P(\bold D|\bold \theta, \mathcal M) \cdot P(\bold \theta | \mathcal M)}{P(\bold D | \mathcal M)} = \frac{\mathcal{L}(\theta)\cdot\pi(\theta)}{\mathcal{Z}}.
\end{equation}
In this expression, $\bld \theta$ denotes the parameters of the model $\mathcal M$ that we aim to fit, and $\bld D$ represents the observed data points \citep{siviaDataAnalysisBayesian2006}. The term $P(\bld \theta | \mathcal M)$, also written as $\pi(\bld \theta)$, is the prior distribution, which encodes our initial beliefs about the parameter probability distribution. The likelihood, $P(\bld D|\bld \theta, \mathcal M)$ or $\mathcal L(\bld \theta)$, quantifies the probability of the data under the assumption that the model and parameters are correct. The posterior distribution, $P(\bld \theta|\bld D, \mathcal M)$ or $\mathcal P(\bld \theta)$, represents the updated probability of the parameters after accounting for the data. Finally, $P(\bld D|\mathcal M)$, known as the Bayesian evidence or $\mathcal Z$, serves as a normalization constant which can be used for model comparison. Due to the linearity of the equations it is also possible to solve for the parameters using a least squares solver \citep{roqueReceiverDesignREACH2025}, although we do not do so in this work.

We use a Gaussian likelihood for calibration \citep{roqueBayesianNoiseWave2021} of the form,
\begin{equation} \label{e:gaussian_likelihood}
    \mathcal L = \frac{1}{\sqrt{2\pi \sigma^2}}\exp\left\{-\frac{1}{2\sigma^2} (\bold{T}_s - \bold{X\Theta})^T (\bold{T}_s - \bold{X\Theta}) \right\},
\end{equation}
where $\sigma$ is the calibrator noise parameter. Since we are using $\Gamma$-weighted temperatures which normalise the noise on the temperatures \citep{kirkhamAccountingNoiseSingularities2025} we are justified in using this single parameter for all calibrators.

\subsection{Surface Fitting}\label{s:methods-surface}

\begin{figure}
    \centering
    \includegraphics[width=\linewidth]{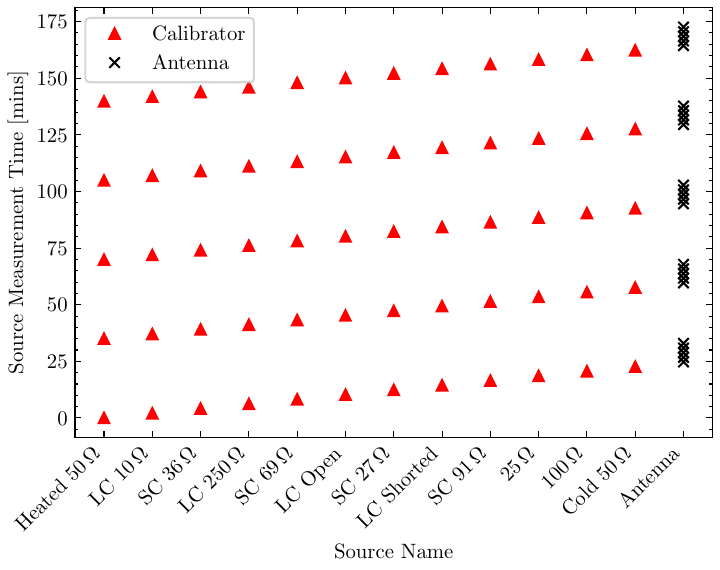}
    \caption{An example of a REACH observation strategy showing the order that the twelve calibration sources (red triangles) are measured in. In between calibration observations the antenna (black crosses) is measured five times. Here LC and SC refer to sources connected to the long $10\,$m cable and the short $2\,$m cable respectively.}
    \label{fig:observation-strat}
\end{figure}
\begin{figure*}
    \centering
    \includegraphics[width=0.7\linewidth]{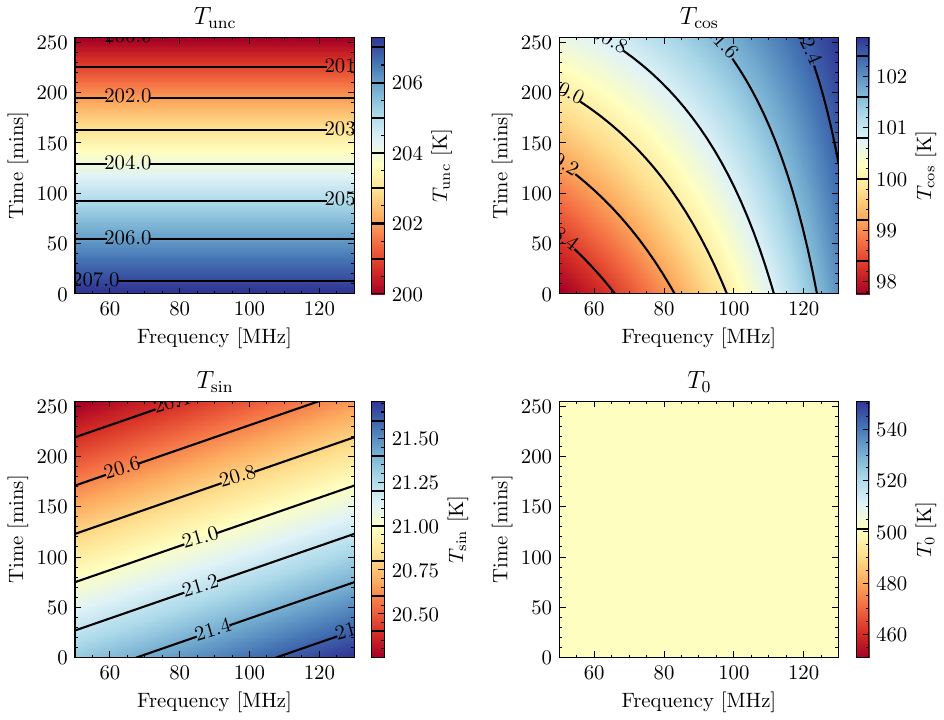}
    \caption{Surface plots of the true noise wave parameters which were used to generate the simulated calibration dataset. $T_\mathrm{unc}$ is second order in time and zeroth order in frequency, $T_\mathrm{cos}$ is first order in time and second order in frequency, $T_\mathrm{sin}$ is first order in time and first order in frequency, and $T_0$ is zeroth order in both time and frequency.}
    \label{fig:true-noise-waves}
\end{figure*} 

Previous REACH noise wave calibration methods \citep{roqueBayesianNoiseWave2021, roqueReceiverDesignREACH2025, kirkhamAccountingNoiseSingularities2025} do not explicitly consider the time information when calibrating the receiver\footnote{Another time-based calibration method can be seen in \cite{leeneyRadiometerCalibrationUsing2025a} which adds time as an input to the calibration neural network.}. This can be problematic since the antenna is measured in between measurements of the calibrators and so, should the state of the LNA change significantly during an observation, we would require an interpolation of the noise wave parameters to accurately calibrate the antenna. The times at which sources are measured in a example REACH observation is plotted in figure \ref{fig:observation-strat} where the twelve calibrators are measured then the antenna spectra is measured five times. 

In this paper we present an extension of the conjugate priors method \citep{roqueBayesianNoiseWave2021} which fits a polynomial surface in frequency and in time to the noise wave parameters in order to extend the solutions to the times at which the antenna is measured. Previously with the polynomial methods we were fitting 1-dimensional frequency-dependent polynomials,
\begin{equation}
    T_\mathrm{NWP}(\nu) = \sum_i a_i\cdot\nu^i,
\end{equation}
to each of the noise wave parameters. However we can extend this to include the time-dependence by fitting a frequency-time surface
\begin{equation}
    T_\mathrm{NWP}(\nu, t) = \sum_i \sum_j A_{ij}\cdot \nu^i t^j.
\end{equation}
We can then rewrite the noise wave parameter for the $i$th calibrator measured at time $t_i$ as a vector across frequency,
\begin{equation}
    \bold{T}_{\mathrm{NWP},i} = \bold{B}_{\mathrm{NWP},i}\bold{\Theta}_\mathrm{NWP},
\end{equation}
where
\begin{equation}
    \bold{\Theta}_\mathrm{NWP} = \begin{pmatrix}
        A_{00}\\
        A_{01}\\
        A_{10}\\
        A_{11}\\
        A_{02}\\
        \vdots
    \end{pmatrix},
\end{equation}
and
\begin{equation}
    \bold{B}_{\mathrm{NWP},i} = \begin{pmatrix}
1 & t_i & \nu_0 & \nu_0 t_i & t_i^2 & \dots \\
1 & t_i & \nu_1 & \nu_1 t_i & t_i^2 & \dots \\
1 & t_i  & \nu_2 & \nu_2 t_i & t_i^2 & \dots \\
\vdots & \vdots & \vdots & \vdots & \vdots & \ddots
\end{pmatrix}.
\end{equation}
Note that there are different $\bold{B}$ matrices for each noise wave parameter as each NWP has its own set of time and frequency polynomial orders.

The calibration equation is then
\begin{align}
\begin{split}
    \bold{T}_\mathrm{s}(t_i) &= \bold{X}_{\mathrm{unc},i}\bold{B}_{\mathrm{unc},i}\bold{\Theta}_\mathrm{unc} + \bold{X}_{\mathrm{cos},i} \bold{B}_{\mathrm{cos},i} \bold{\Theta}_\mathrm{cos} \\
    &+ \bold{X}_{\mathrm{sin},i} \bold{B}_{\mathrm{sin},i} \bold{\Theta}_\mathrm{sin} + \bold{X}_{\mathrm{NS},i} \bold{B}_{\mathrm{NS},i} \bold{\Theta}_\mathrm{NS}\\
    &+ \bold{X}_{\mathrm{L},i} \bold{B}_{\mathrm{L},i} \bold{\Theta}_\mathrm{L},
\end{split}
\end{align}
which can be written as
\begin{equation}
    \bold{T}_\mathrm{s}(t_i) = \bold{X}_i\bold{\Theta},
\end{equation}
where
\begin{equation}
    \bold{X}_i = \begin{pmatrix}
        \bold{X}_{\mathrm{unc},i} \bold{B}_{\mathrm{unc},i} \\ \bold{X}_{\mathrm{cos},i}\bold{B}_{\mathrm{cos},i} \\ \bold{X}_{\mathrm{sin},i}\bold{B}_{\mathrm{sin},i} \\ \bold{X}_{\mathrm{NS},i}\bold{B}_{\mathrm{NS},i} \\ \bold{X}_{\mathrm{L},i}\bold{B}_{\mathrm{L},i}
    \end{pmatrix}^T,
\end{equation}
and
\begin{equation}
    \bold{\Theta} = \begin{pmatrix}
        \bold{\Theta}_\mathrm{unc} \\
        \bold{\Theta}_\mathrm{cos} \\
        \bold{\Theta}_\mathrm{sin} \\
        \bold{\Theta}_\mathrm{NS} \\
        \bold{\Theta}_\mathrm{L}
    \end{pmatrix}.
\end{equation}
We can stack the calibrators to give
\begin{equation}
    \bold{T}_\mathrm{s} = \begin{pmatrix}
        \bold{T}_{\mathrm{s},1}\\
        \bold{T}_{\mathrm{s},2}\\
        \vdots
    \end{pmatrix},
\end{equation}
and
\begin{equation}
    \bold{X} = \begin{pmatrix}
        \bold{X}_1\\
        \bold{X}_2\\
        \vdots
    \end{pmatrix}.
\end{equation}

The posterior distributions of the polynomial surface parameters, $\bld{\Theta}$, are then sampled using the conjugate priors method of \cite{roqueBayesianNoiseWave2021}. To jointly determine the polynomial orders of the surface in both time and frequency, we perform a gradient ascent on the Bayesian evidence surface to determine the orders with the highest evidence. Since the evidence incorporates an `Occam penalty' \citep{hergtBayesianEvidenceTensortoscalar2021}, this mitigates overfitting of the data, improving our interpolation in time of the noise wave parameters. We note that using a polynomial surface makes the assumption that the noise wave parameters are smooth in both frequency and time so it may be necessary to explore other basis functions.




%
%
%
%


\subsection{Simulating Data}\label{s:methods-simulation}

\begin{figure*}
    \centering
    \includegraphics[width=0.9\linewidth]{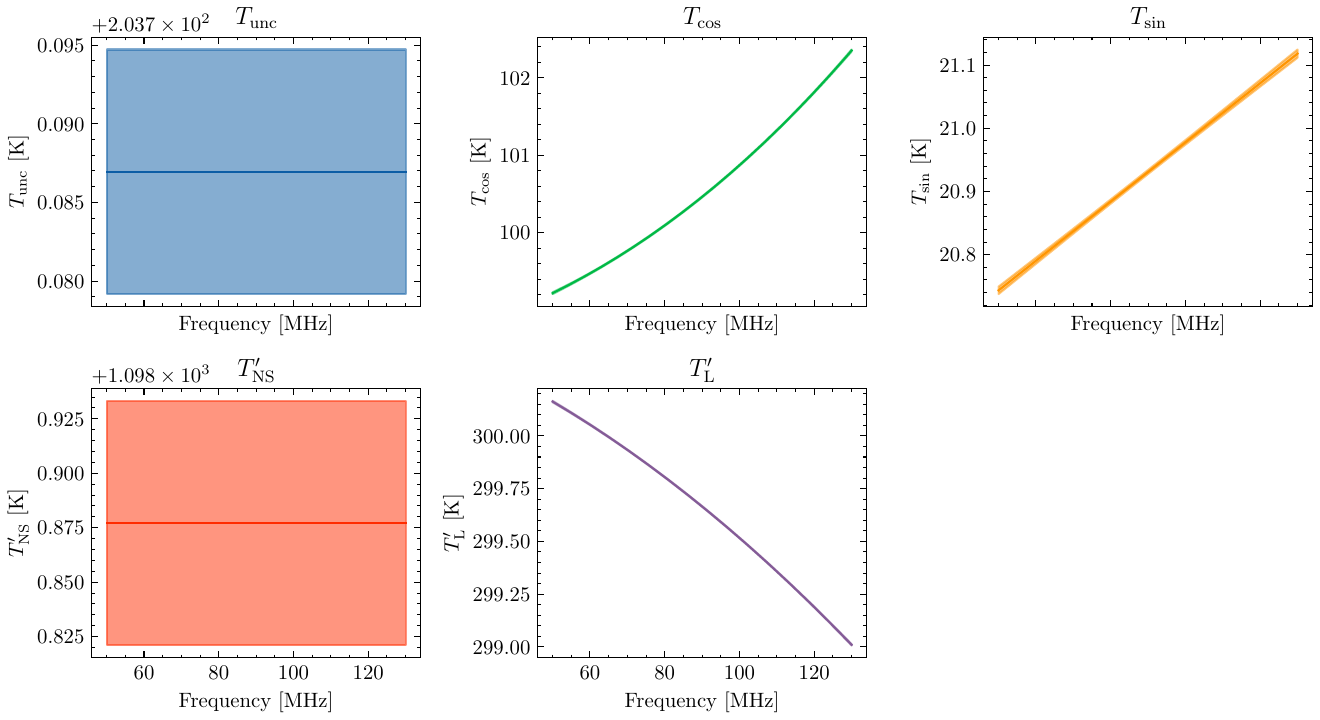}
    \caption{One-dimensional function posterior plots of the polynomial fits to the simulated calibration dataset. The posterior mean is plotted as a line and the shaded area shows $\pm1\sigma$ around the mean. As this method does not account for time variation of the noise wave parameters these solutions are effectively time-averaged calibration solutions.}
    \label{fig:one-D_param_solutions}
\end{figure*}

To test these methods we generated a simulated dataset using reflection coefficient measurements from the REACH instrument on site. To simulate a drifting LNA we modelled the noise wave parameters, $T_\mathrm{unc}$, $T_\mathrm{cos}$ and $T_\mathrm{sin}$, as low-order polynomial surfaces. The receiver noise offset, $T_0$, was modelled as constant in time and frequency. The measured power spectral density of a calibrator can then be calculated to be
\begin{align}
\begin{split}
    P_s = kBg (&T_s (1 - |\Gamma_s|^2)|F_s|^2+ T_\mathrm{unc} |\Gamma_s|^2|F_s|^2\\
    &+ T_\mathrm{cos} \mathrm{Re}(\Gamma_s F_s)  + T_\mathrm{sin} \mathrm{Im}(\Gamma_s F_s) + T_0),
\end{split}
\end{align}
where $k$ is the Boltzmann constant, $B$ is the frequency channel width and $g$ is the source-independent gain of the LNA \citep{rogersAbsoluteCalibrationWideband2012}. We set $g$ to the magnitude of the forward S-parameter of the REACH LNA, $|S_{21}|$. The four simulated polynomial surfaces for the noise wave parameters, $T_\mathrm{unc}$, $T_\mathrm{cos}$, $T_\mathrm{sin}$ and $T_0$ are plotted in figure \ref{fig:true-noise-waves}.

As a stand-in for an antenna, we simulate a validation source with a flat $5000\,$K spectrum across frequency, the noise source at $1400\,K$ and the hot load at $370\,$K while all other sources are at an ambient temperature of $300\,$K. We do not include the effects of cable temperature gradients in this simulation.

The integration time is $t_\mathrm{int} = 60\,$s and we add radiometric noise to the PSDs with a standard deviation of
\begin{equation}
    \sigma_{P_s} = \epsilon\frac{P_s}{\sqrt{t_\mathrm{int}B}},
\end{equation}
where $\epsilon = \frac{1}{100}$ is a factor which is chosen to be an arbitrarily small value to show the residual structure in the final calibrated temperatures.

Preliminary tests on REACH data from the instrument on site show that any time variation of the noise wave parameters is below the radiometric noise level of the data, so the surface fitting algorithm finds surfaces which are zeroth-order in time to be optimal. As this is mathematically equivalent to fitting the REACH data with the \cite{roqueBayesianNoiseWave2021} 1-dimensional polynomials, we choose not to include this data here. This may suggest that the LNA variation in REACH is long-term, highlighting the utility of this method for combining many days' observations to help better characterise the noise wave parameters.

\section{Results}\label{s:results}

\begin{figure*}
    \centering
    \begin{subfigure}{0.49\textwidth}
        \centering
        \includegraphics[width=\linewidth]{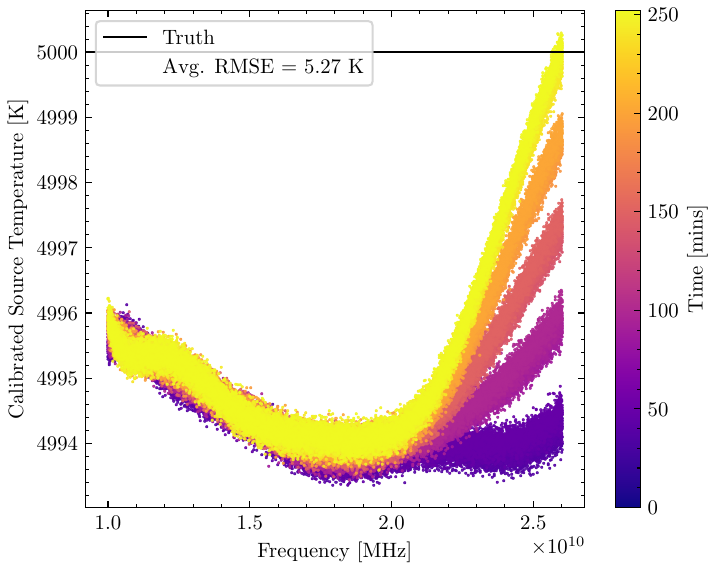}
        \caption{One-dimensional polynomial fit with the \cite{roqueBayesianNoiseWave2021} calibration equation, equation \ref{e:calibration_equation}.}
        \label{fig:one-D_antenna_sol}
    \end{subfigure}\hfill
    \begin{subfigure}{0.49\textwidth}
        \centering
        \includegraphics[width=\linewidth]{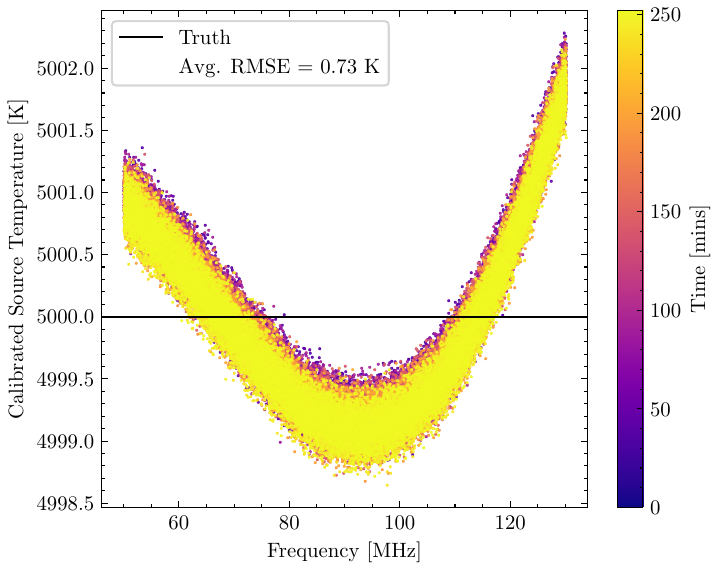}
        \caption{Surface fitting with the \cite{roqueBayesianNoiseWave2021} calibration equation, equation \ref{e:calibration_equation}.}
        \label{fig:surface_fit_antenna_sol}
    \end{subfigure}\hfill
    \begin{subfigure}{0.49\textwidth}
        \centering
        \includegraphics[width=\linewidth]{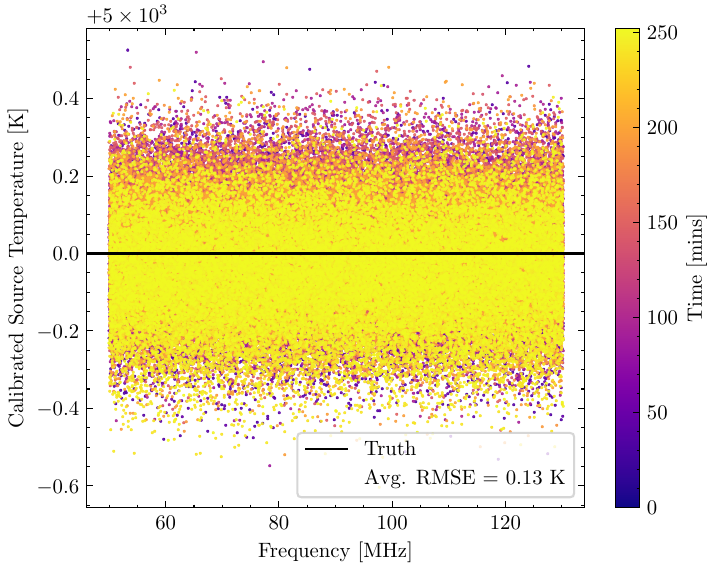}
        \caption{Surface fitting with this work's calibration equation, equation \ref{e:calibration_eqn_TNSTL}, removing the assumptions on the cold load's and noise source's reflection coefficients.}
        \label{fig:surface_fit_TNSTL_antenna_sol}
    \end{subfigure}\hfill
    \caption{Plots of the calibrated validation source solutions for the three methods described, with the true temperature plotted as a black horizontal line. It can be seen that introducing the surface fitting method removes the drift in the solution, and further using the calibration equation in equation \ref{e:calibration_eqn_TNSTL} removes the chromatic residual.}
\end{figure*}

\begin{table*}
    \centering
    \begin{tabular}{ccccc}
         \textbf{Fit Function}&  \textbf{Calibration Equation}& \textbf{Drift in Solution?} & \textbf{Chromatic Residual?} & \textbf{Validation Source RMSE [K]} \\\hline\hline
         1-D Polynomial& \cite{roqueBayesianNoiseWave2021}, equation \ref{e:calibration_equation} & {\color{red}Yes} & {\color{red}Yes} & 5.27\\\hline
         Polynomial Surface&\cite{roqueBayesianNoiseWave2021}, equation \ref{e:calibration_equation}& {\color{ForestGreen}No} & {\color{red}Yes}& 0.73 \\\hline
         Polynomial Surface& This work's, equation \ref{e:calibration_eqn_TNSTL} & {\color{ForestGreen}No} & {\color{ForestGreen}No} & 0.13  \\
    \end{tabular}
    \caption{Results of running the three methods on the simulated data. For each we indicate the fit function used, the calibration equation used (equation \ref{e:calibration_equation} which has reflection coefficient assumptions or equation \ref{e:calibration_eqn_TNSTL} which removes those assumptions), whether the calibrated solution has a drift over time, whether the calibrated solution has a visible chromatic residual, and finally the time-average of the root mean square error of the calibrated solution.}
    \label{tab:fit_results}
\end{table*}

\subsection{1-Dimensional Polynomial Fit}

The results of calibrating the simulated data with the 1-D dimensional polynomial fit described in \cite{roqueBayesianNoiseWave2021} are shown in figure \ref{fig:one-D_param_solutions}. Here the functional posteriors of the five noise wave parameters are plotted, with the solid line representing the posterior mean and the shaded area indicating $\pm1\sigma$. This method does not take in any time information, using the five observations of the twelve calibrators effectively as sixty calibration sources. As a result this leads to a time averaged solution to the calibration equations, ignoring any drift that may be happening.

Figure \ref{fig:one-D_antenna_sol} shows the resulting calibrated validation source temperature as a result of this 1-D method, with the true $5000\,$K temperature shown by the black horizontal line. A summary of all results can be seen in table \ref{tab:fit_results}. Taking the time average of the root mean squared errors (RMSE), we see that there is a large error of $5.27\,$K. Here we see a large chromatic residual to the true temperature which drifts with time. We also see a significant monochromatic absolute offset. This highlights the need to take into account the time information as fairly small changes in the noise wave parameters over time can have dramatic impacts on your final calibrated temperature.

\subsection{Surface Fitting}

\begin{figure*}
    \centering
    \includegraphics[width=0.9\linewidth]{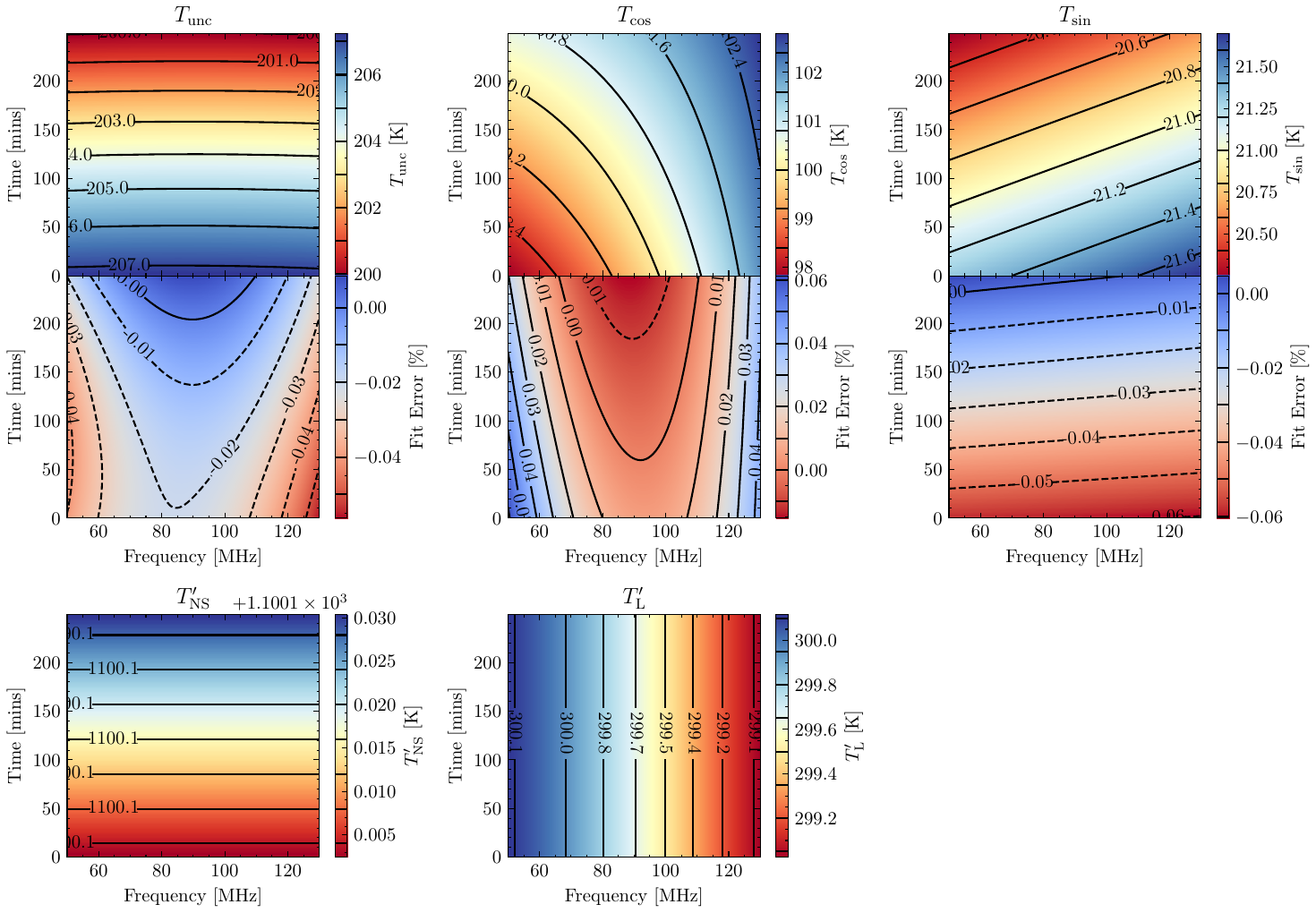}
    \caption{Plots of the fitted calibration solutions for the polynomial surface fits to the \protect\cite{roqueBayesianNoiseWave2021} calibration equation (equation \ref{e:simplified_calibration_eqn}). The fit error compared with the ground truth has been plotted for $T_\mathrm{unc}$, $T_\mathrm{cos}$ and $T_\mathrm{sin}$. There is no fit error for $T'_\mathrm{NS}$ and $T'_\mathrm{L}$ as these are effective temperatures. The surfaces here have been resampled onto equally spaced times to get a smooth plot.}
    \label{fig:surface_fit_param_solutions}
\end{figure*}


Next, we fitted the simulated data with a polynomial surface using the calibration equation from \cite{roqueBayesianNoiseWave2021}. The posterior means of the five fitted parameters are shown in figure \ref{fig:surface_fit_param_solutions}, together with the difference between this method's fits and the ground truth. Since $T_\mathrm{NS}'$ and $T_\mathrm{L}'$ are not parameters set in the simulation, we do not have a value for the fit error. Looking at the errors on $T_\mathrm{unc}$, $T_\mathrm{cos}$ and $T_\mathrm{sin}$, we see that this method performs well with, at worst, a 0.06\% fit error.

We can see the calibrated validation source solution for this method in figure \ref{fig:surface_fit_antenna_sol}. While the solution drift has mostly been corrected for by the surface fitting, there is still a significant chromatic calibration residual visible. Despite this, we see a large improvement over the 1-dimensional fit, with a smaller absolute offset from the truth and the time-averaged RMSE dropping by 86\% to 0.73 K.

The degeneracies highlighted in section \ref{s:methods-assumptions} may be to blame for the chromatic residual in the calibrated temperature. We can see evidence of this in the fit to $T_\mathrm{NS}'$ in figure \ref{fig:surface_fit_param_solutions} as we see that the fitted value changes by $0.03\,$K over time. Since the noise source temperature and all reflection coefficients are constant over time in the simulation, this time variation in the fit must be resulting from the $T_\mathrm{unc}$, $T_\mathrm{cos}$ and $T_\mathrm{sin}$ terms bleeding in to $T_\mathrm{NS}'$. This degeneracy is hence problematic and could have biased our fitting of the parameters, resulting in the imperfect calibration we see here.

\subsection{Surface Fitting with This Work's Calibration Equation}

\begin{figure*}
    \centering
    \includegraphics[width=0.7\linewidth]{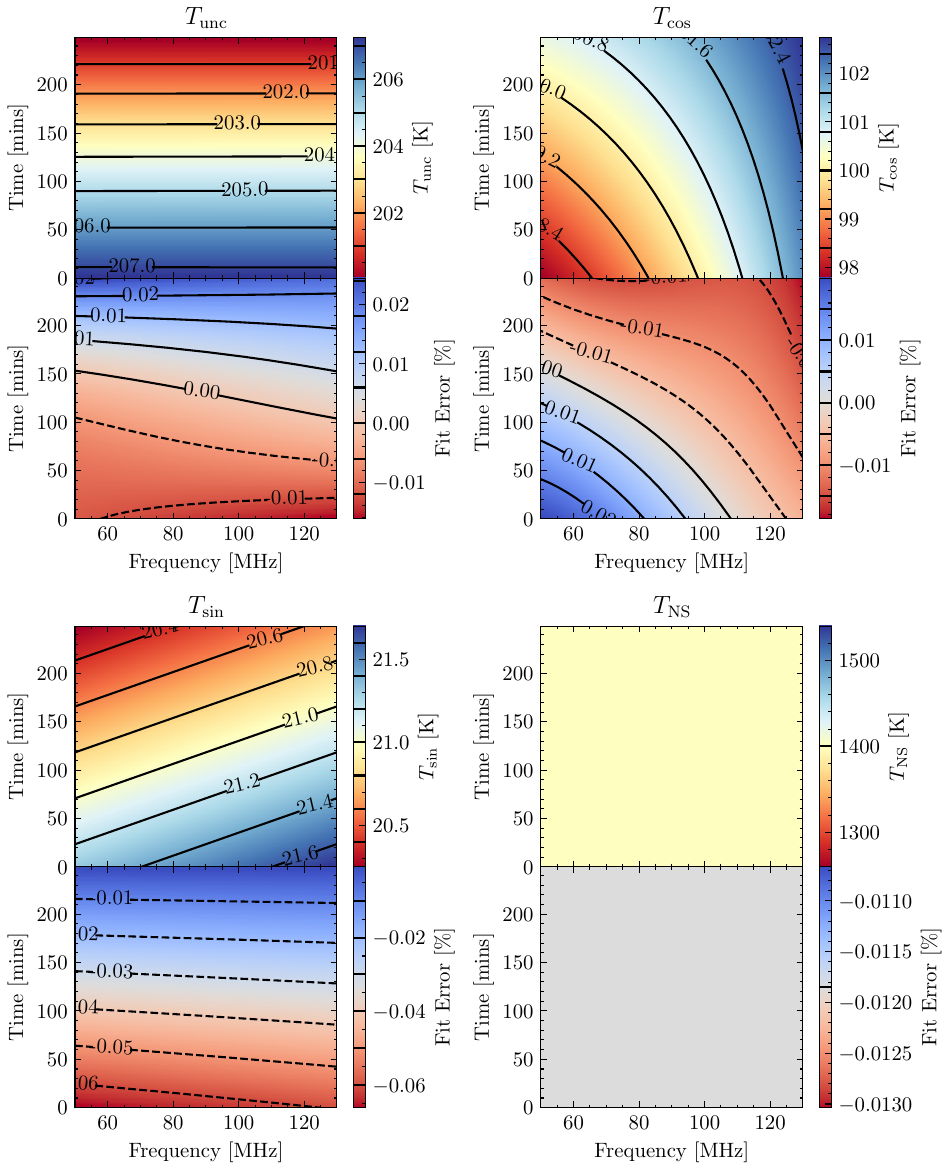}
    \caption{Plots of the fitted calibration solutions for the polynomial surface fits to this work's calibration equation (equation \ref{e:calibration_eqn_TNSTL}). The fit error compared with the ground truth has been plotted for the four parameters. The surfaces here have been resampled onto equally spaced times to get a smooth plot.}
    \label{fig:surface_fit_TNSTL_param_solutions}
\end{figure*}


Finally, we used the calibration equation from equation \ref{e:calibration_eqn_TNSTL} to fit the simulated calibration dataset. As this has explicit terms for the noise source and cold load reflection coefficients and uses the measured value of the cold load temperature, this should remove the degeneracies we encountered with the \cite{roqueBayesianNoiseWave2021} calibration equation. Figure \ref{fig:surface_fit_TNSTL_param_solutions} shows the posterior means of the solutions to the four parameters and the difference between this method's fits and the ground truth. In comparison to the \cite{roqueBayesianNoiseWave2021} calibration equation, there is no large improvement with the $T_\mathrm{sin}$ fit but the fits to $T_\mathrm{unc}$ and $T_\mathrm{cos}$ have improved, with a six times reduction in error for $T_\mathrm{cos}$.

Similarly we see the improvement in the calibrated validation source temperature in figure \ref{fig:surface_fit_TNSTL_antenna_sol}. There is now no longer any drift in calibration solution, nor any chromatic residual visible in the spectra. Comparing this with the previous surface fitting method, we see a further 82\% reduction in the time-averaged RMSE to $0.13\,$K. Overall, this method shows a 97\% improvement in the time-averaged RMSE compared with the original \cite{roqueBayesianNoiseWave2021} 1-dimensional calibration method.

The improvement in the fit quality can be attributed to the removal of degeneracies from the calibration equation fit parameters. We see that in figure \ref{fig:surface_fit_TNSTL_param_solutions} the fit to $T_\mathrm{NS}$ is now a flat plane at the simulated noise source temperature, as we would expect. Any remaining absolute offset can be attributed to the 0.01\% error in the fit to $T_\mathrm{NS}$, the parameter which sets the absolute scale of the calibrated temperatures. If the true noise source temperature is known to sufficient accuracy then this can be inserted in the calibration equation, removing the need to fit for $T_\mathrm{NS}$ and accurately defining the absolute temperature scale.

\section{Conclusions}\label{s:conclusions}

In this paper we introduced a new method to extend previous REACH calibration methods to include time variation in the noise wave parameters. To do so we fit the noise wave parameters with polynomial surfaces and use the previous conjugate priors framework from \cite{roqueBayesianNoiseWave2021} to quickly evaluate the parameter posterior distributions. We also highlight the degeneracies in the noise wave parameters and present a reformulation of the calibration equation to remove these. To test these methods we simulated a REACH system with polynomial surface noise wave parameters to emulate a drifting receiver system.

Calibrating a simulated $5000\,$K flat validation source without accounting for system drifting resulted in a large chromatic residual which changes over time. Using the surface fitting method with the \cite{roqueBayesianNoiseWave2021} calibration equation results in a calibrated temperature solution which has had the drift corrected but still has a significant chromatic residual. Furthermore there is evidence of the degeneracies in the fits to $T_\mathrm{NS}'$ and $T_\mathrm{L}'$ which are biasing the fit, resulting in an imperfect calibration. Finally, using the surface fitting method with the less degenerate calibration equation introduced in equation \ref{e:calibration_eqn_TNSTL} results in a flat, static calibrated temperature solution. While the fit to $T_\mathrm{sin}$ has not improved, we find that the error in the recovery of $T_\mathrm{unc}$ and $T_\mathrm{cos}$ has improved up to six times as a result of removing the degeneracies from the parameter fits. In total, introducing the surface fitting and removing the reflection coefficient assumptions resulted in a 97\% reduction in RMSE of the residual to the validation source.

The primary limitation of this method is that it still carries the assumption that the system (gain, $g$, and receiver noise, $T_0$) is constant over the course of the Dicke switching from source to cold source to noise source. Since this assumption is baked into equations \ref{e:full_calibration_eqn} and \ref{e:calibration_eqn_TNSTL}, it is not trivial to remove this assumption. Further improvement to the method could be made by changing the set of basis functions used to fit the noise wave parameters. In this method, we used polynomials as the basis functions but it may be necessary to consider other functions, such as the Fourier basis or 2-D Gaussian processes, to properly capture the frequency and time behaviour of the LNA.


\section*{Acknowledgements}

CJK was supported by Science and Technology Facilities Council grant number ST/V506606/1. DJA was supported by Science and Technology Facilities Council grant number ST/X00239X/1. EdLA was supported by Science and Technology Facilities Council grant number ST/V004425/1. We would also like to thank the Kavli Foundation for their support of REACH.

\section*{Data Availability}

The data that supported the findings of this article will be shared on reasonable request to the corresponding author.



\bibliographystyle{mnras}
\bibliography{21cm_Cosmology} 

\begin{thebibliography}{}
\makeatletter
\relax
\def\mn@urlcharsother{\let\do\@makeother \do\$\do\&\do\#\do\^\do\_\do\%\do\~}
\def\mn@doi{\begingroup\mn@urlcharsother \@ifnextchar [ {\mn@doi@} {\mn@doi@[]}}
\def\mn@doi@[#1]#2{\def\@tempa{#1}\ifx\@tempa\@empty \href {http://dx.doi.org/#2} {doi:#2}\else \href {http://dx.doi.org/#2} {#1}\fi \endgroup}
\def\mn@eprint#1#2{\mn@eprint@#1:#2::\@nil}
\def\mn@eprint@arXiv#1{\href {http://arxiv.org/abs/#1} {{\tt arXiv:#1}}}
\def\mn@eprint@dblp#1{\href {http://dblp.uni-trier.de/rec/bibtex/#1.xml} {dblp:#1}}
\def\mn@eprint@#1:#2:#3:#4\@nil{\def\@tempa {#1}\def\@tempb {#2}\def\@tempc {#3}\ifx \@tempc \@empty \let \@tempc \@tempb \let \@tempb \@tempa \fi \ifx \@tempb \@empty \def\@tempb {arXiv}\fi \@ifundefined {mn@eprint@\@tempb}{\@tempb:\@tempc}{\expandafter \expandafter \csname mn@eprint@\@tempb\endcsname \expandafter{\@tempc}}}

\bibitem[\protect\citeauthoryear{Barkana}{Barkana}{2018}]{barkanaPossibleInteractionBaryons2018}
Barkana R.,  2018, \mn@doi [Nature] {10.1038/nature25791}, 555, 71

\bibitem[\protect\citeauthoryear{Bevins, Fialkov, {de Lera Acedo}, Handley, Singh, Subrahmanyan  \& Barkana}{Bevins et~al.}{2022}]{bevinsAstrophysicalConstraintsSARAS2022}
Bevins H. T.~J.,  Fialkov A.,  {de Lera Acedo} E.,  Handley W.~J.,  Singh S.,  Subrahmanyan R.,   Barkana R.,  2022, \mn@doi [Nat Astron] {10.1038/s41550-022-01825-6}, 6, 1473

\bibitem[\protect\citeauthoryear{Bowman, Rogers  \& Hewitt}{Bowman et~al.}{2008}]{bowmanEmpiricalConstraintsGlobal2008}
Bowman J.~D.,  Rogers A. E.~E.,   Hewitt J.~N.,  2008, \mn@doi [The Astrophysical Journal] {10.1086/528675}, 676, 1

\bibitem[\protect\citeauthoryear{Bowman, Rogers, Monsalve, Mozdzen  \& Mahesh}{Bowman et~al.}{2018}]{bowmanAbsorptionProfileCentred2018}
Bowman J.~D.,  Rogers A. E.~E.,  Monsalve R.~A.,  Mozdzen T.~J.,   Mahesh N.,  2018, \mn@doi [Nature] {10.1038/nature25792}, 555, 67

\bibitem[\protect\citeauthoryear{Cumner et~al.,}{Cumner et~al.}{2022}]{cumnerRadioAntennaDesign2022}
Cumner J.,  et~al., 2022, \mn@doi [J. Astron. Instrum.] {10.1142/S2251171722500015}, 11, 2250001

\bibitem[\protect\citeauthoryear{Dhandha et~al.,}{Dhandha et~al.}{2025}]{dhandhaNarrowingDiscoverySpace2025}
Dhandha J.,  et~al., 2025, Narrowing the Discovery Space of the Cosmological 21-Cm Signal Using Multi-Wavelength Constraints, \mn@doi{10.48550/arXiv.2508.13761}

\bibitem[\protect\citeauthoryear{Dicke}{Dicke}{1946}]{dickeMeasurementThermalRadiation1946}
Dicke R.~H.,  1946, \mn@doi [Review of Scientific Instruments] {10.1063/1.1770483}, 17, 268

\bibitem[\protect\citeauthoryear{Feng \& Holder}{Feng \& Holder}{2018}]{fengEnhancedGlobalSignal2018}
Feng C.,  Holder G.,  2018, \mn@doi [ApJL] {10.3847/2041-8213/aac0fe}, 858, L17

\bibitem[\protect\citeauthoryear{Furlanetto, Peng~Oh  \& Briggs}{Furlanetto et~al.}{2006}]{furlanettoCosmologyLowFrequencies2006}
Furlanetto S.~R.,  Peng~Oh S.,   Briggs F.~H.,  2006, \mn@doi [Physics Reports] {10.1016/j.physrep.2006.08.002}, 433, 181

\bibitem[\protect\citeauthoryear{Hergt, Handley, Hobson  \& Lasenby}{Hergt et~al.}{2021}]{hergtBayesianEvidenceTensortoscalar2021}
Hergt L.~T.,  Handley W.~J.,  Hobson M.~P.,   Lasenby A.~N.,  2021, \mn@doi [Phys. Rev. D] {10.1103/PhysRevD.103.123511}, 103, 123511

\bibitem[\protect\citeauthoryear{Hills, Kulkarni, Meerburg  \& Puchwein}{Hills et~al.}{2018}]{hillsConcernsModellingEDGES2018}
Hills R.,  Kulkarni G.,  Meerburg P.~D.,   Puchwein E.,  2018, \mn@doi [Nature] {10.1038/s41586-018-0796-5}, 564, E32

\bibitem[\protect\citeauthoryear{Kirkham et~al.,}{Kirkham et~al.}{2025}]{kirkhamAccountingNoiseSingularities2025}
Kirkham C.~J.,  et~al., 2025, Accounting for {{Noise}} and {{Singularities}} in {{Bayesian Calibration Methods}} for {{Global}} 21-Cm {{Cosmology Experiments}} (\mn@eprint {arXiv} {2412.14023}), \mn@doi{10.48550/arXiv.2412.14023}

\bibitem[\protect\citeauthoryear{Leeney et~al.,}{Leeney et~al.}{2025}]{leeneyRadiometerCalibrationUsing2025a}
Leeney S. A.~K.,  et~al., 2025, Radiometer {{Calibration}} Using {{Machine Learning}}, \mn@doi{10.48550/arXiv.2504.16791}

\bibitem[\protect\citeauthoryear{Meys}{Meys}{1978}]{meysWaveApproachNoise1978}
Meys R.,  1978, \mn@doi [IEEE Transactions on Microwave Theory and Techniques] {10.1109/TMTT.1978.1129303}, 26, 34

\bibitem[\protect\citeauthoryear{Monsalve, Rogers, Bowman  \& Mozdzen}{Monsalve et~al.}{2017}]{monsalveCALIBRATIONEDGESHIGHBAND2017}
Monsalve R.~A.,  Rogers A. E.~E.,  Bowman J.~D.,   Mozdzen T.~J.,  2017, \mn@doi [ApJ] {10.3847/1538-4357/835/1/49}, 835, 49

\bibitem[\protect\citeauthoryear{Monsalve, Greig, Bowman, Mesinger, Rogers, Mozdzen, Kern  \& Mahesh}{Monsalve et~al.}{2018}]{monsalveResultsEDGESHighband2018}
Monsalve R.~A.,  Greig B.,  Bowman J.~D.,  Mesinger A.,  Rogers A. E.~E.,  Mozdzen T.~J.,  Kern N.~S.,   Mahesh N.,  2018, \mn@doi [ApJ] {10.3847/1538-4357/aace54}, 863, 11

\bibitem[\protect\citeauthoryear{Monsalve, Fialkov, Bowman, Rogers, Mozdzen, Cohen, Barkana  \& Mahesh}{Monsalve et~al.}{2019}]{monsalveResultsEDGESHighBand2019}
Monsalve R.~A.,  Fialkov A.,  Bowman J.~D.,  Rogers A. E.~E.,  Mozdzen T.~J.,  Cohen A.,  Barkana R.,   Mahesh N.,  2019, \mn@doi [The Astrophysical Journal] {10.3847/1538-4357/ab07be}, 875, 67

\bibitem[\protect\citeauthoryear{Monsalve et~al.,}{Monsalve et~al.}{2023}]{monsalveMapperIGMSpin2023}
Monsalve R.~A.,  et~al., 2023, Mapper of the {{IGM Spin Temperature}} ({{MIST}}): {{Instrument Overview}} (\mn@eprint {arXiv} {2309.02996}), \mn@doi{10.48550/arXiv.2309.02996}

\bibitem[\protect\citeauthoryear{Philip et~al.,}{Philip et~al.}{2019}]{philipProbingRadioIntensity2019}
Philip L.,  et~al., 2019, \mn@doi [J. Astron. Instrum.] {10.1142/S2251171719500041}, 08, 1950004

\bibitem[\protect\citeauthoryear{Price et~al.,}{Price et~al.}{2018}]{priceDesignCharacterizationLargeaperture2018}
Price D.~C.,  et~al., 2018, \mn@doi [Monthly Notices of the Royal Astronomical Society] {10.1093/mnras/sty1244}, 478, 4193

\bibitem[\protect\citeauthoryear{Price, Tong, Sutinjo, Greenhill  \& Patra}{Price et~al.}{2023}]{priceMeasuringNoiseParameters2023}
Price D.~C.,  Tong C.-Y.~E.,  Sutinjo A.~T.,  Greenhill L.~J.,   Patra N.,  2023, \mn@doi [IEEE Transactions on Microwave Theory Techniques] {10.1109/TMTT.2022.3225317}, 71, 1102

\bibitem[\protect\citeauthoryear{Rogers \& Bowman}{Rogers \& Bowman}{2012}]{rogersAbsoluteCalibrationWideband2012}
Rogers A. E.~E.,  Bowman J.~D.,  2012, \mn@doi [Radio Science] {10.1029/2011RS004962}, 47

\bibitem[\protect\citeauthoryear{Roque, Handley  \& {Razavi-Ghods}}{Roque et~al.}{2021}]{roqueBayesianNoiseWave2021}
Roque I. L.~V.,  Handley W.~J.,   {Razavi-Ghods} N.,  2021, \mn@doi [Monthly Notices of the Royal Astronomical Society] {10.1093/mnras/stab1453}, 505, 2638

\bibitem[\protect\citeauthoryear{Roque et~al.,}{Roque et~al.}{2025}]{roqueReceiverDesignREACH2025}
Roque I. L.~V.,  et~al., 2025, \mn@doi [Exp Astron] {10.1007/s10686-024-09975-3}, 59, 7

\bibitem[\protect\citeauthoryear{Shaver, Windhorst, Madau  \& {de Bruyn}}{Shaver et~al.}{1999}]{shaverCanReionizationEpoch1999}
Shaver P.~A.,  Windhorst R.~A.,  Madau P.,   {de Bruyn} A.~G.,  1999, \mn@doi [Astronomy and Astrophysics] {10.48550/arXiv.astro-ph/9901320}, 345, 380

\bibitem[\protect\citeauthoryear{Sims \& Pober}{Sims \& Pober}{2020}]{simsTestingCalibrationSystematics2020}
Sims P.~H.,  Pober J.~C.,  2020, \mn@doi [Monthly Notices of the Royal Astronomical Society] {10.1093/mnras/stz3388}, 492, 22

\bibitem[\protect\citeauthoryear{Singh, Subrahmanyan, Shankar, Rao, Girish, Raghunathan, Somashekar  \& Srivani}{Singh et~al.}{2018}]{singhSARASSpectralRadiometer2018}
Singh S.,  Subrahmanyan R.,  Shankar N.~U.,  Rao M.~S.,  Girish B.~S.,  Raghunathan A.,  Somashekar R.,   Srivani K.~S.,  2018, \mn@doi [Exp Astron] {10.1007/s10686-018-9584-3}, 45, 269

\bibitem[\protect\citeauthoryear{Singh et~al.,}{Singh et~al.}{2022}]{singhDetectionCosmicDawn2022}
Singh S.,  et~al., 2022, \mn@doi [Nat Astron] {10.1038/s41550-022-01610-5}, 6, 607

\bibitem[\protect\citeauthoryear{Sivia}{Sivia}{2006}]{siviaDataAnalysisBayesian2006}
Sivia D.~S.,  2006, Data Analysis: A {{Bayesian}} Tutorial., 2nd ed. / d.s. sivia with j. skilling. edn.
Oxford Science Publications, University Press, Oxford

\bibitem[\protect\citeauthoryear{{de Lera Acedo} et~al.,}{{de Lera Acedo} et~al.}{2022}]{deleraacedoREACHRadiometerDetecting2022a}
{de Lera Acedo} E.,  et~al., 2022, \mn@doi [Nat Astron] {10.1038/s41550-022-01709-9}, 6, 984

\makeatother
\end{thebibliography}



\bsp	
\label{lastpage}
\end{document}